# Title: An Adaptive Soft Plasmonic Nanosheet Resonator

Xinghua Wang,[†] Tanju Yildirim,[†] Kae Jye Si,[†] Ankur Sharma, Yunzhou Xue, Qinghua Qin, Qiaoliang Bao, Wenlong Cheng,[*] and Yuerui Lu[*]

Dr. X. Wang
Research School of Engineering, College of Engineering and Computer Science, the Australian National University, Canberra, ACT, 2601, Australia
Unmanned Systems Research Center, National Innovation Institute of Defense Technology, Academy of Military Science, Beijing, 100166, China
A. Sharma, Q Qin, Prof. Y. Lu
Research School of Engineering, College of Engineering and Computer Science, the Australian National University, Canberra, ACT, 2601, Australia
E-mail: yuerui.lu@anu.edu.au
Dr. T. Yildirim, Dr. Yunzhou Xue
College of Chemistry and Environmental Engineering, Shenzhen University, Shenzhen 518060, P. R. China
K. J. Si, Prof. W. Cheng
Department of Chemical Engineering, Faculty of Engineering, Monash University, Victoria, 3800, Australia
E-mail: wenlong.cheng@monash.edu
Prof. Q. Bao
Department of Materials Science and Engineering, ARC Centre of Excellence in Future Low-Energy Electronics Technologies (FLEET), Monash University, Clayton, Victoria, Australia
[†] These authors equally contributed to this work

Current micro/nanomechanical system are usually based on rigid crystalline semiconductors that normally have high quality factors but lack adaptive responses to variable frequencies – a capability ubiquitous for communications in the biological world, such as bat and whale calls. Here, we demonstrate a soft mechanical resonator based on a freestanding organic-inorganic hybrid plasmonic superlattice nanosheet, which can respond adaptively to either incident light intensity or wavelength. This is achieved because of strong plasmonic coupling in closely-packed nanocrystals which can efficiently concentrate and convert photons into heat. The heat causes the polymer matrix to expand, **leading to a change in the nanomechanical properties of the plasmonic nanosheet. Notably, the adaptive frequency responses are also reversible and the responsive ranges are fine-tunable by adjusting the constituent nanocrystal building blocks. We believe that our plasmonic nanosheets may open a new route to design next-generation intelligent bio-mimicking opto-mechanical resonance systems.**

Through millennia of evolutionary advances, plants and animals throughout nature have designed unique mechanisms for locomotion and interaction with the rapidly changing environment.[1] A large basis of these evolutionary mechanisms involve biological components that emit variable frequency signals that are used for deterring predators, sensing, defense and attracting partners for breeding.[2] For instance, bats normally change their outgoing navigation signal for echolocation.[3] Bats can generate variable frequency signals over wideband spectra for ambient flight conditions giving them a wide range of view. After detecting potential prey, this frequency increases over finite distances for precise predation (**Figure 1a**). Moths use a tympanic membrane where the pretension can be increased by expansion of the ear cavity as a defense mechanism against bats.[4] Dolphins are able to use clicking to generate sonar signals over large spectral ranges for high bandwidth communications.[5] Mammalian cochlea are an efficient source of transforming sound frequencies within the auditory nerve having little power requirements.[6] From these examples prevalent throughout nature, tunable frequency devices are an integral part for communications and bandwidth, sensing the surrounding environment, interaction between complex systems and they also provide great insight into evolutionary adaptations.[7] For animals, these variable frequency mechanisms are relatively quite small in size, have simple tuning mechanism designs and low tuning power requirements. Furthermore, plants, animals and viruses are highly adapted to using sunlight (photothermal) for energy and cellular interaction.[8]

Nanoelectromechanical systems (NEMS) have become a promising technology due to their low mass, high stiffness and robust behaviour.[9] NEMS technology has become more predominant in daily life as manufacturing techniques for the synthesis of nano-scale devices and their systematic integration with electronic components and circuitry has advanced.[10] In particular, their low mass, high stiffness and quality factor make NEMS well suited as resonating devices in high frequency spectra; this allows NEMS resonators to be widely adopted in many potential science and engineering applications, such as, vibration and optical absorbers, electronics, energy-harvesters and as biomolecular sensors.[11-13] Resonator design is based on matching a devices fundamental frequency with the surrounding environment. However, this leads to an inherent drawback, if the operating frequency of a device does not match the intended use or the ambient frequency shifts, then the resonator does not operate; moreover, manufacturing imperfections also alter the intended resonance frequency of a device which can lead to frequency variability. Due to these reasons, adaptive devices are of significant interest because they are frequency tunable due to a controllable external mechanism. Furthermore, when integrating with micro and nano-scale devices, the mechanism for an adaptive response also needs to be considered, as this will affect the size, manufacturing procedure, tunable range, design complexity, cost and feasibility of a proposed design. Compared with other tuning techniques, optical stimulation allows for a non-contact method using an external light source.[14] A major concern for optically tunable devices is the achievable broadband frequency range and the power required to tune the devices. In general, organic optically tuned resonators inhibit large tunability, and the power required to tune the device is high, in contrast, inorganic resonators usually have low tuning power requirements but their tunable range is narrow (Figure 1b).[14-18]

Herein, we show that low tuning power and high resonance tunability can be simultaneously achieved with our ligand-mediated plasmonic nanosheet resonators (Figure 1b). The nanosheets are fabricated using polystyrene-capped *Au@Ag* nanocubes in a drying-mediated self-assembly process.[19-21] The light absorbed by these nanoparticles is transformed into heat due to the photo-thermal effect. The nanoparticles are embedded in a freestanding super-lattice to enhance the induced optical-thermal-mechanical coupling.[22] This type of hybrid organic-inorganic freestanding nanosheet can be used as a nano-electromechanical resonator with extremely high stiffness and low mass for high bandwidth sensing and actuating applications.[23-24] The additional advantage of the proposed device is that it requires a non-contact tuning method that does not require leads, electrodes or other electrical components integrated within the device, allowing for easier stimulation, high efficiency and easier manufacturing. Compared with traditional substrate supported nanoparticle assemblies in the literature, the authors use this design to demonstrate two major findings in this work. The first is that by using an organic-inorganic plasmonic nanosheet, a wideband tunable range from 55.2 to 34.1 kHz can be achieved for relatively low light intensities. Moreover, the second major finding of this work is that using a plasmonic nanosheet enables wavelength selective photon sensing and actuation, this behaviour is significantly increased under plasmonic resonance conditions. Due to the advanced near field light-matter interaction inside the plasmonic nanosheet, the resonance frequency is adaptive to either incident light intensity or incident wavelength which gives the proposed device a substantially robust and adaptive response. This type of technology can also be combined with low-dimensional materials for future opto-mechanical devices.[25-32]

**Results**

Recently, noble metallic nanoparticles have attracted extensive attention due to their extraordinary optical and electrical characteristics,[33-34] these metallic nanoparticles can enhance light absorption and scattering in materials and devices.[35-37] However, absorption and subsequent temperature increasing have often been considered as undesirable effects,[38-39] and most research focuses on the optical properties of metallic nanoparticles.[40-41] Moreover, it has been demonstrated that plasmonic metallic nanoparticles subjected to optical stimulation are an efficient nano heat-source,[42-44] generating localized heating of particles at the nano-scale. This advanced optical-thermal coupling behaviour opens a broad new range of applications, including photo-thermal imaging, cancer targeting and radiation detection.[45-47]

The structure adopted in this work utilizes gold nanoparticles encapsulated in silver cubes, which are embedded within a polystyrene matrix and suspended on a silicon substrate as shown in **Figure 2a**. At the center of the device is a squared shaped freestanding membrane with equal lengths $L$ that can vibrate in the out of plane direction. More details about the fabrication procedure are shown in Supplementary Figure 1 and the corresponding supplementary information. The particles are well aligned inside the polystyrene matrix which act as intermolecular spring linkages. Polystyrene was selected as the matrix because it has strong near field light matter interaction with gold nanoparticle heaters (see Supplementary Figure 2). Characterization was carried out with atomic force microscopy (AFM) and transmission electron microscopy (TEM), and the structural parameters of this nanosheet can be extracted from the obtained high-resolution images (see Figures 2b and c). The nanoparticle diameter is 11 nm, the silver cube has equal lengths of 25 nm and the plasmonic nanosheet has overall thickness of $40 \pm 2$ nm (which is the combination of metal particles embedded in the polystyrene matrix).

Upon successful device fabrication, the dynamic response of the nano-resonator in the frequency-domain was conducted using an external piezoelectric excitation system and the experimental system is schematically illustrated in **Figure 3a**. The dominant resonance peak was observed at approximately 41 kHz and this corresponded to the first fundamental out of plane vibration mode. The experimental frequency-amplitude and frequency-phase curves for a plasmonic nanosheet resonator with side lengths of 300 μm is shown in Figure 3b, subjected to a driving voltage of $\delta V_g = 20$ mV in air. A Lorentzian line fitted as a damped harmonic model yielded a resonance frequency of $f_0 = 41.2$ kHz, and a quality factor value of $Q = 190$. The authors have further characterized the resonance frequency versus $1/L$ for nine different resonators with side lengths ranging from 90 to 300 μm, all resonators were fabricated on a common base. From the fitted red line in Figure 3c, the dependence of the resonance frequency can be extracted as $f_0 \propto 1/L$, which is expected for square shaped membranes.[48-49] The average value of the pretension $T_0$ was experimentally obtained as 0.074 N/m, deviations from the fitted line are due to random built-in pre-tension generated during the fabrication procedure from the heterogeneous contraction at the air-water interface during the drying-mediated self-assembly process. Another source of this dispersion is due to the variation of non-uniformly distributed nanoparticles within the super-lattice between devices. The standing out of plane 2D map of the fundamental mode was experimentally obtained and is shown in Figure 3d. During the mapping process, the excitation frequency was fixed at the resonance frequency and the interferometer laser gun was controlled by a stage controller for lateral movement with a sub-0.5 μm resolution and a grid was setup in order to obtain the experimental mode shape. Furthermore, COMSOL simulations of the fundamental mode shape are shown in Figure 3e, the simulated natural frequency was 42.8 kHz showing good agreement between the experiment and simulation. Moreover, comparing Figures 3d and e, it has been shown that the mapping procedure shows a near similar shape to the simulated mode shape verifying that it is the fundamental mode shape of the resonator.

The resonance frequency of the device is highly sensitive to the internal in-plane tension and this offers the possibility for a tunable resonance frequency. The pre-tension value depends on the fabrication process, however, using the innovative hybrid organic-inorganic configuration proposed in this work, the strong coupling of the light-matter interaction inside the plasmonic nanosheet enables an adaptive response. The frequency-response curves of a plasmonic nanosheet resonator with $L = 250$ μm under different light intensities has been experimentally obtained and shown in **Figure 4a**. The intensity of the incident light used was 0, 5, 10 and 15 mW/cm$^2$ which was measured using a power-meter. For three of the frequency-response curves recorded, there are slight increases and decreases and this is possibly due to interactions with higher modes of the plasmonic resonator. Typically, tuning of the resonator was quite fast using an optical source demonstrating the low power requirements and this was due to the low mass and strong optical-mechanical coupling of the device. It was observed when the plasmonic resonator was subjected to a 15 mW/cm$^2$ intensity, the resonance frequency exhibited an extremely large downward resonance frequency shift from 55.2 to 34.1 kHz (approximately 40% of the devices own fundamental frequency). With increasing light intensity, a consistent down-shift of the devices fundamental frequency was observed and this was due to the polystyrene molecular interlinks axially compressing within the freestanding plasmonic super-lattice due to the hybrid organic-inorganic nature as shown in Figure 4b. This phenomenon was caused by the strong optical-thermal interaction of the plasmonic nanosheet and subsequently this caused the resonance frequency to decrease with increasing light intensity; this behaviour was also enhanced as the two-dimensional super-lattice is single-particle thick (40±2 *nm*) and its resonance frequency is dominated by the in-plane tension of the membrane. During the interaction with an optical source, the electric field strongly drives mobile carriers inside the nanoparticles, and subsequently, the energy gained by the carriers turn into heat through electron-phonon-interaction. Afterwards, the generated heat in the nanoparticles diffuse into the polystyrene leading to an elevated temperature.[50-51] The temperature increase caused by the photo-thermal effect consequently drives internal compression (mechanical softening), shifting the resonance frequency downwards. The authors have demonstrated the optically adaptive response of the hybrid organic-inorganic plasmonic nanosheet has both high tunability and low tuning power requirements. To further verify the added advantages of the organic-inorganic design proposed in this work, controlled experiments for the frequency-shift properties between a plasmonic super-lattice nanosheet and a *Si$_3$N$_4$* membrane (which is a conventional nano-resonator in the literature) with the same side lengths were performed, and the experimentally obtained results are shown in Figure 4c. It was observed that the resonance frequency of the $Si_3N_4$ membrane resonator remains constant regardless of the incident light intensity. In contrast, using the plasmonic nanosheet in this work, the nanoparticles absorb the optical energy and react with the light and this drives a temperature differential in the polystyrene, leading to axial in-plane compression of the polystyrene molecular links. Since the resonance frequency of the super-lattice membrane is dependent upon the internal in-plane tension, a shift in the devices temperature will lead to optically adaptive frequencies.[52] 2D COMSOL simulations of the electric field, temperature distribution and stress around a single gold-silver nanocube particle are shown in Figures 4d, e and f, respectively (the nanoparticle diameter is 11 nm and the silver cube has equal lengths of 25 nm). It can be seen that the electric field and temperature distribution is uniform with the wave moving equally outwards in both directions; however, for the stress distribution, the polystyrene develops large internal stress from the photothermal excitation particularly around the points of each nanocube and in turn this can change the intrinsic in-plane tension of the plasmonic nanosheet. The advantages of the photothermal excitation are there are no physical contacts (i.e. electrodes or wiring), high temperature differentials, high efficiency and stability for low irradiation (i.e. reversibility). Detailed modelling of the nanoparticle heating effect, shifting the resonance frequency of the device and the temperature frequency shift are given in the Supplementary Information. For the device in this work, it was found that large exposure of light for significant periods of time can cause color changes in the plasmonic superlattice, however, if the irradiation is low the resonator cools back down to its initial temperature and has the same resonance frequency.

Plasmonic structures increase internal heat generation, have low absorption and scatter light over wide frequency ranges, these key attributes allow plasmonic structures to be used for spectroscopy and enhancing solar cell efficiency.[53-54] In this work, the authors demonstrate selective photon sensing using the plasmonic effect, the heating effect becomes especially strong under plasmonic resonance conditions. When the energy of the incident photons is close to the plasmonic resonance frequency of the nanoparticles, the internal heat generation is larger and this is highly dependent upon the morphology of the metallic nanoparticles. According to the measured extinction spectra in **Figure 5a**, the plasmonic resonance frequency of the super-lattice nanosheet is ideally located in the visible and near-ultraviolet regions (which is around 490 nm). The resonance frequency shift strongly depends on the wavelength of the incident light, this means that incident light of relatively weak intensity can tune the resonance frequency easily, in contrast, using higher wavelengths of light requires more power for tuning. The main reason for this behaviour is due to strong optical absorption within the vicinity of the plasmonic resonance. The frequency-shift behaviour of a plasmonic nanosheet resonator using two different incident light wavelengths at various light intensities has also been experimentally investigated and is shown in Figure 5b. It was observed that when using a wavelength of 500 nm (close to the plasmonic resonance), the fundamental frequency shifted downwards 1456 Hz when subjected to a 2 mW/cm$^2$ light intensity. However, using the same light intensity and changing the incident wavelength to 750 nm, a 700 Hz downshift was observed. Therefore, the optically adaptive behaviour of the device presented in this work can be selectively tuned using either the incident light intensity or incident wavelength allowing for greater operational flexibility in practical applications. The strong optical-thermal-mechanical coupling in this work enables adaptive tuning of the device using low optical power and is well suited for the micro and nano domains. Having dual conditions for wideband behaviour allows for greater control of sensors and actuators with ultra-high sensitivity, which makes this type of resonator well suited for telecommunications, circuits and precision instrumentation.

**Conclusion and Outlook**

**In conclusion, self-assembled two-dimensional plasmonic nanosheet resonators with optically adaptive wideband resonance frequencies have been demonstrated. This hybrid**

organic-inorganic resonator is a new type of broadband nanomechanical system integrating plasmonic, photothermal and mechanical properties into a nanometer-thick free-standing nanosheet. The mechanical response of the resonators with and without incident light was experimentally investigated and it was observed that the plasmonic effect down-shifted the resonance frequency of a device and enabled an optically adaptive frequency tunability over 21 kHz with very low incident light intensities. The adaptive photothermal mechanical resonating effect demonstrated here in conjunction with the extremely low mass, and large surface area of the nanosheets makes our wideband tunable plasmonic nanosheet resonator well-suited for high bandwidth applications. In addition, their frequency-shift tuning properties show that our soft resonator device could be used for selective photon sensing and actuating applications. The above attributes indicate that our soft plasmonic nanosheet resonators may enable the design of future intelligent bio-mimetic adaptive nanophotonics and nanomechanics.

**Experimental Section**

*Fabrication of the free-standing plasmonic nanosheet:* The freestanding plasmonic nanosheet was self-fabricated using a drying mediated technique and was deposited onto a silicon substrate with square holes, resulting in the formation of a freestanding plasmonic resonator. The silicon substrate with square holes was fabricated by micro-machining processes, including silicon nitride thin film deposition, lithography, nitride plasma etching, silicon KOH etching, etc. The square shaped resonator has equal lengths *L* and has a single particle thickness in the out of plane direction of approximately 40 nm. The plasmonic nanosheet resonator is considered to be fully clamped around the square shaped holes and resonates in the out of plane direction as confirmed through experiments and simulations (see Figures 3d and e).

*Mechanical resonance frequency characterization:* A piezoelectric plate was mounted on the backside of the silicon die for actuation, and the vibration of the membrane was detected with an optical interferometer and lock-in amplifier. The input signal to the piezoelectric actuation system was controlled *via* a signal generator.

*Simulations:* Simulations for the resonance frequency of the plasmonic nanosheet resonator were conducted using COMSOL Multiphysics 5.1. The membrane module was selected for the interface physics to account for the in-plane tension dominated natural frequency of the plasmonic resonator. Table S1 was used for the input parameters for the simulations. A mesh of 5856 elements using a triangular distribution was used and the eigenfrequency physics solver was used for determining the resonance frequency of the square shaped plasmonic nanosheet resonator. For the 2D simulation in Figure 4, the electromagnetics and heat transfer in solids physics modules were used. A 0.001 convergence criterion was used for all simulations in this manuscript.

**Supporting Information**
Supporting Information is available from the Wiley Online Library or from the author.

**Acknowledgements**
We would like to acknowledge financial support from the ANU PhD scholarship, the China Research Council PhD scholarship, National Science Foundation China (No. 61775147), Australian Research Council (No. DP180103238), and the ANU Major Equipment Committee (17MEC25, 14MEC34). Received: ((will be filled in by the editorial staff)) Revised: ((will be filled in by the editorial staff)) Published online: ((will be filled in by the editorial staff))

**References**

[1]   B. McInroe, H. C. Astley, C. Gong, S. M. Kawano, P. E. Schiebel, J. M. Rieser, H. Choset, R. W. Blob, D. I. Goldman, *Science* **2016**, *353*, 154.
[2]   J.-X. Shen, Z.-M. Xu, Z.-L. Yu, S. Wang, D.-Z. Zheng, S.-C. Fan, *Nature Communications* **2011**, *2*, 342.
[3]   S. Greif, B. M. Siemers, *Nature Communications* **2010**, *1*, 107.
[4]   J. Guerreiro, J. C. Jackson, J. F. Windmill, "Bio-inspired frequency agile acoustic system", presented at *SENSORS, 2016 IEEE,* 2016.
[5]   C. Capus, Y. Pailhas, K. Brown, D. M. Lane, P. W. Moore, D. Houser, *The Journal of the Acoustical Society of America* **2007**, *121*, 594.
[6]   S. Mandal, S. M. Zhak, R. Sarpeshkar, *IEEE J. Solid-State Circuits* **2009**, *44*, 1814.
[7]   N. H. Fletcher, *The Journal of the Acoustical Society of America* **2004**, *115*, 2334.
[8]   S. Hong, M. Y. Lee, A. O. Jackson, L. P. Lee, *Light: Science &Amp; Applications* **2015**, *4*, e267.
[9]   G. Luo, Z.-Z. Zhang, G.-W. Deng, H.-O. Li, G. Cao, M. Xiao, G.-C. Guo, L. Tian, G.-P. Guo, *Nature Communications* **2018**, *9*, 383.
[10]  D. Akinwande, N. Petrone, J. Hone, *Nature Communications* **2014**, *5*, 5678.
[11]  A. Savchenkov, A. Matsko, V. Ilchenko, D. Seidel, L. Maleki, *Opt. Lett.* **2011**, *36*, 3338.

[12] J. Thompson, B. Zwickl, A. Jayich, F. Marquardt, S. Girvin, J. Harris, *Nature* **2008**, *452*, 72.
[13] Y. Lu, S. Peng, D. Luo, A. Lal, *Nature communications* **2011**, *2*, 578.
[14] V. Sathi, N. Ehteshami, J. Nourinia, *Org. Electron.* **2012**, *13*, 1192.
[15] N. Ehteshami, V. Sathi, *J. Electron. Mater.* **2013**, *42*, 162.
[16] T. Larsen, S. Schmid, L. G. Villanueva, A. Boisen, *ACS Nano* **2013**, *7*, 6188.
[17] D. Kim, E. Lee, M. Cho, C. Kim, Y. Park, T. Kouh, *Appl. Phys. Lett.* **2013**, *102*, 203502.
[18] O. Hajime, K. Takehito, O. Koji, M. Imran, Y. Hiroshi, *Applied Physics Express* **2009**, *2*, 062202.
[19] K. J. Si, D. Sikdar, Y. Chen, F. Eftekhari, Z. Xu, Y. Tang, W. Xiong, P. Guo, S. Zhang, Y. Lu, Q. Bao, W. Zhu, M. Premaratne, W. Cheng, *ACS Nano* **2014**, *8*, 11086.
[20] K. J. Si, Y. Chen, W. Cheng, *Mater. Today* **2016**, *19*, 363.
[21] C. Wenlong, *EPL (Europhysics Letters)* **2017**, *119*, 48004.
[22] X. Wang, K. J. Si, J. Yang, X. Wu, Q. Qin, W. Cheng, Y. Lu, "Ultra-sensitive photon sensor based on self-assembled nanoparticle plasmonic membrane resonator", presented at *Micro Electro Mechanical Systems (MEMS), 2016 IEEE 29th International Conference on*, 2016.
[23] A. Castellanos-Gomez, M. Poot, G. A. Steele, H. S. J. van der Zant, N. Agraït, G. Rubio-Bollinger, *Adv. Mater.* **2012**, *24*, 772.
[24] C. Chen, S. Rosenblatt, K. I. Bolotin, W. Kalb, P. Kim, I. Kymissis, H. L. Stormer, T. F. Heinz, J. Hone, *Nat. Nanotechnol.* **2009**, *4*, 861.
[25] L. Zhang, H. Yan, X. Sun, M. Dong, T. Yildirim, B. Wang, B. Wen, G. P. Neupane, A. Sharma, Y. Zhu, *Nanoscale* **2019**, *11*, 418.
[26] L. Zhang, A. Sharma, Y. Zhu, Y. Zhang, B. Wang, M. Dong, H. T. Nguyen, Z. Wang, B. Wen, Y. Cao, *Adv. Mater.* **2018**, *30*, 1803986.
[27] J. Pei, J. Yang, T. Yildirim, H. Zhang, Y. Lu, *Adv. Mater.* **2019**, *31*, 1706945.
[28] Y. Zhu, Z. Li, L. Zhang, B. Wang, Z. Luo, J. Long, J. Yang, L. Fu, Y. Lu, *ACS Appl. Mater. Interfaces* **2018**, *10*, 43291.
[29] J. Pei, J. Yang, Y. Lu, *IEEE Journal of Selected Topics in Quantum Electronics* **2017**, *23*.
[30] T. Vogl, G. Campbell, B. C. Buchler, Y. Lu, P. K. Lam, *ACS Photonics* **2018**.
[31] A. Sharma, H. Yan, L. Zhang, X. Sun, B. Liu, Y. Lu, *Acc. Chem. Res.* **2018**, *51*, 1164.
[32] S. Peng, A. Lal, D. Luo, Y. Lu, *Nanoscale* **2017**, *9*, 6953.
[33] W. L. Barnes, A. Dereux, T. W. Ebbesen, *nature* **2003**, *424*, 824.
[34] J. A. Fan, C. Wu, K. Bao, J. Bao, R. Bardhan, N. J. Halas, V. N. Manoharan, P. Nordlander, G. Shvets, F. Capasso, *science* **2010**, *328*, 1135.
[35] Z. Nie, A. Petukhova, E. Kumacheva, *Nat. Nanotechnol.* **2010**, *5*, 15.
[36] K. A. Willets, R. P. Van Duyne, *Annu. Rev. Phys. Chem.* **2007**, *58*, 267.
[37] Y. Ito, K. Matsuda, Y. Kanemitsu, *Physical review B* **2007**, *75*, 033309.
[38] G. Baffou, R. Quidant, *Laser Photonics Rev.* **2013**, *7*, 171.
[39] Y.-F. Xiao, Y.-C. Liu, B.-B. Li, Y.-L. Chen, Y. Li, Q. Gong, *Physical Review A* **2012**, *85*, 031805.
[40] Y. Kang, Y. Gong, Z. Hu, Z. Li, Z. Qiu, X. Zhu, P. M. Ajayan, Z. Fang, *Nanoscale* **2015**, *7*, 4482.
[41] J. N. Anker, W. P. Hall, O. Lyandres, N. C. Shah, J. Zhao, R. P. Van Duyne, *Nat. Mater.* **2008**, *7*, 442.
[42] A. O. Govorov, H. H. Richardson, *Nano today* **2007**, *2*, 30.
[43] G. Baffou, R. Quidant, F. J. García de Abajo, *ACS nano* **2010**, *4*, 709.
[44] D. Boyer, P. Tamarat, A. Maali, B. Lounis, M. Orrit, *Science* **2002**, *297*, 1160.
[45] L. Cognet, S. Berciaud, D. Lasne, B. Lounis, ACS Publications, 2008.

[46] E. Y. Lukianova-Hleb, X. Ren, R. R. Sawant, X. Wu, V. P. Torchilin, D. O. Lapotko, *Nat. Med.* **2014**, *20*, 778.
[47] F. Yi, H. Zhu, J. C. Reed, E. Cubukcu, *Nano Lett.* **2013**, *13*, 1638.
[48] A. Castellanos‐Gomez, R. van Leeuwen, M. Buscema, H. S. van der Zant, G. A. Steele, W. J. Venstra, *Adv. Mater.* **2013**, *25*, 6719.
[49] Z. Wang, H. Jia, X. Zheng, R. Yang, Z. Wang, G. Ye, X. Chen, J. Shan, P. X.-L. Feng, *Nanoscale* **2015**, *7*, 877.
[50] H. H. Richardson, Z. N. Hickman, A. O. Govorov, A. C. Thomas, W. Zhang, M. E. Kordesch, *Nano Lett.* **2006**, *6*, 783.
[51] H. H. Richardson, M. T. Carlson, P. J. Tandler, P. Hernandez, A. O. Govorov, *Nano Lett.* **2009**, *9*, 1139.
[52] A. L. Tchebotareva, P. V. Ruijgrok, P. Zijlstra, M. Orrit, *Laser Photonics Rev.* **2010**, *4*, 581.
[53] J. Wang, S. V. Boriskina, H. Wang, B. r. M. Reinhard, *Acs Nano* **2011**, *5*, 6619.
[54] Z. Yue, B. Cai, L. Wang, X. Wang, M. Gu, *Science advances* **2016**, *2*, e1501536.

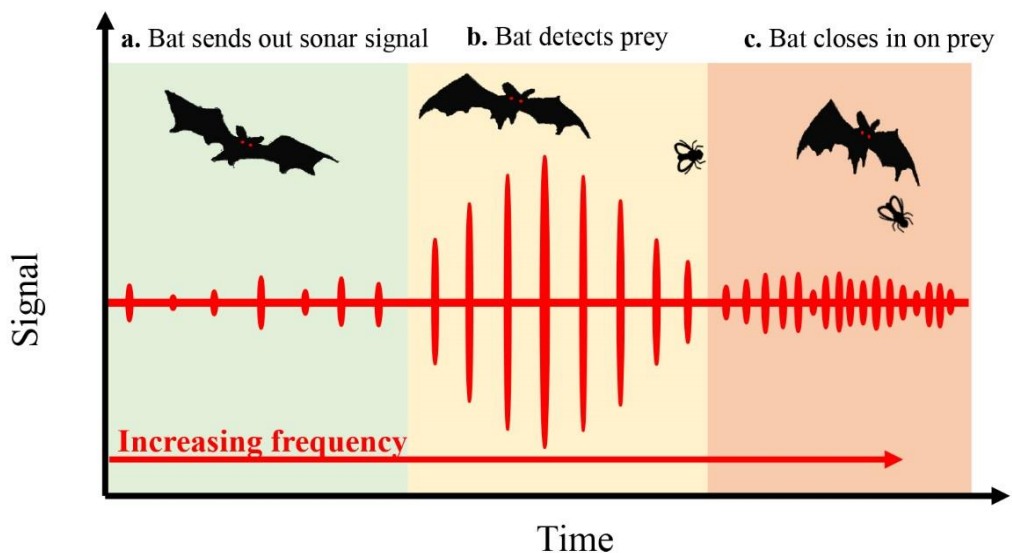

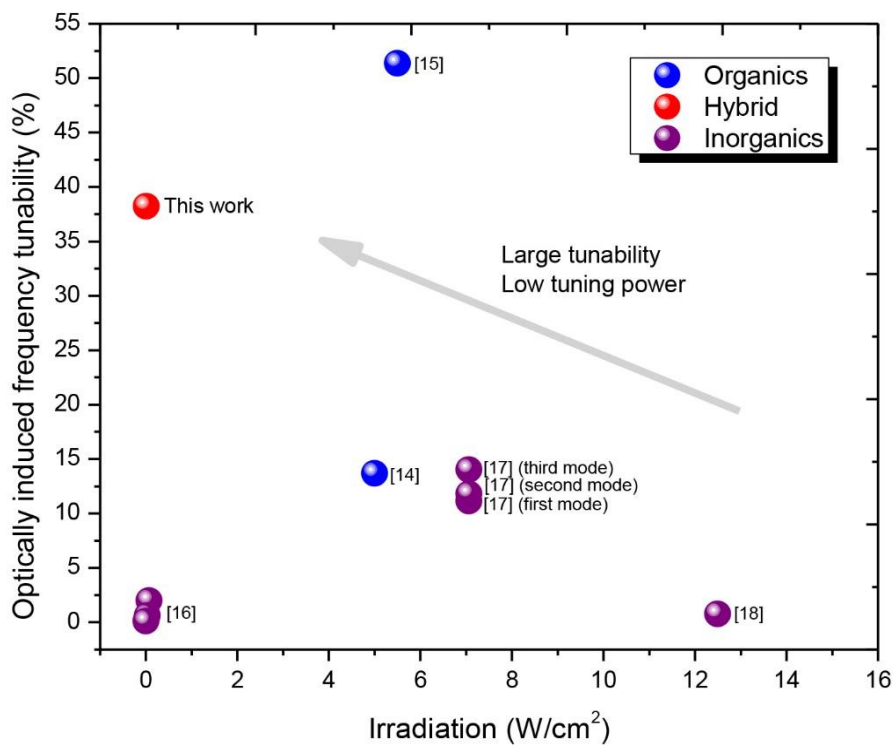

**Figure 1. Motivation for hybrid organic-inorganic design.** a) Frequency tunability of bat calls during predation. b) Comparison of organic, inorganic and hybrid optically tuned devices in the literature, the device presented in this work has both high tuning and low power requirements.

**a**

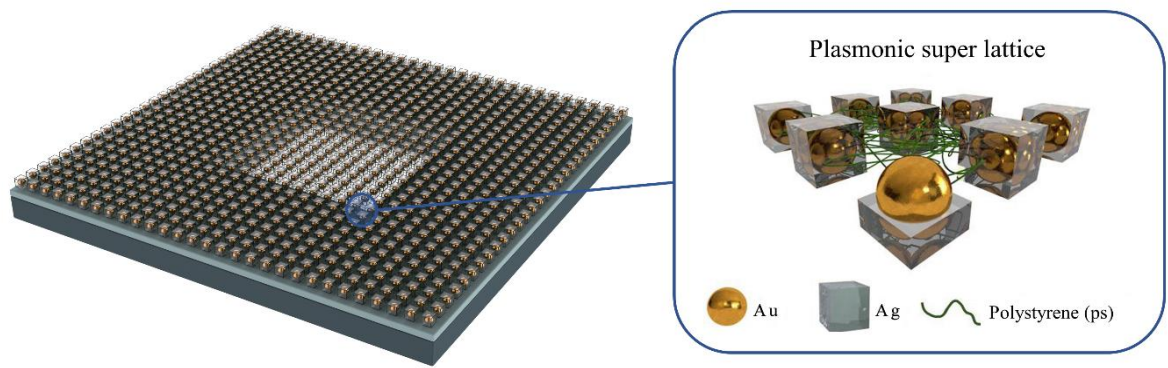

**b**

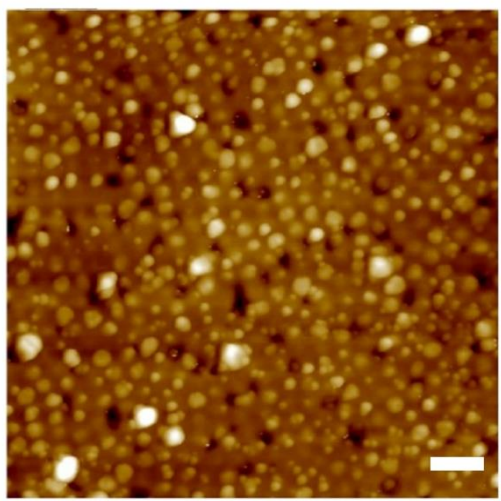

**c**

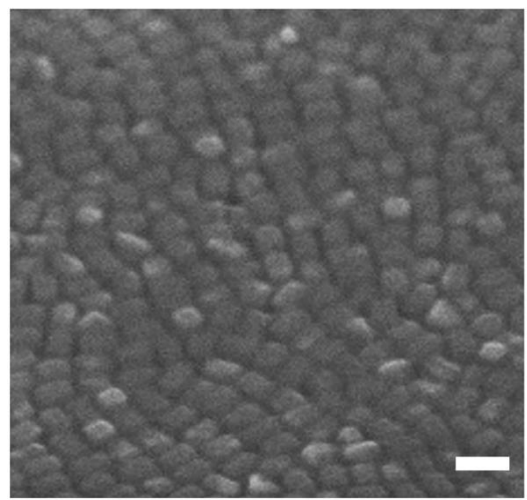

**Figure 2. Configuration of the hybrid organic-inorganic nanoplasmonic resonator.** a) Schematic design of the device structure showing the connection of particles and the center section represents the free standing plasmonic nanosheet membrane that can resonate in the out of plane direction. b) Atomic force microscopy (AFM) image of the nanoplasmonic membrane (scale bar 250 nm). c) Transmission electron microscopy (TEM) image of the nanoplasmonic membrane (scale bar 250 nm)

a

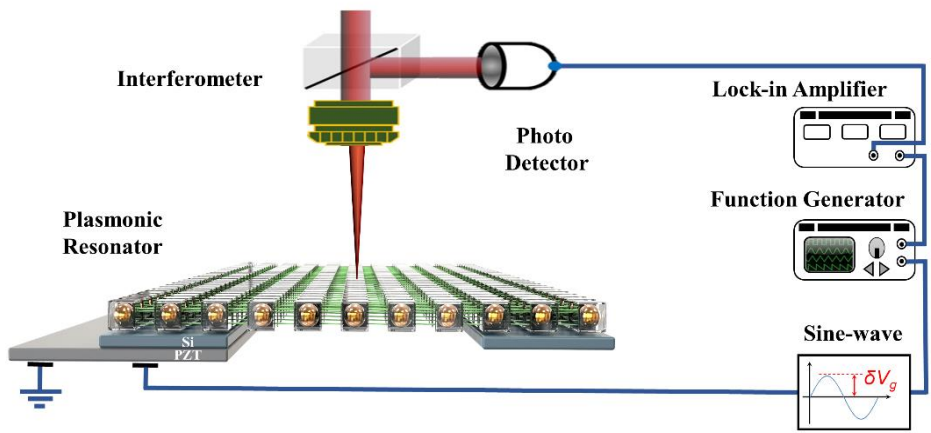

b

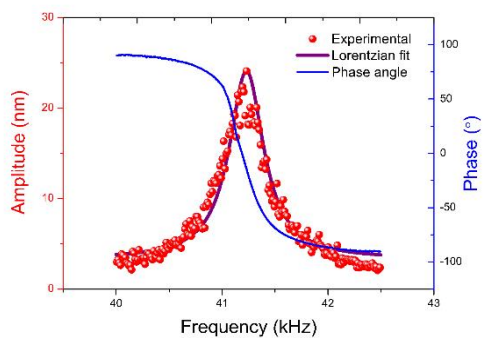

c

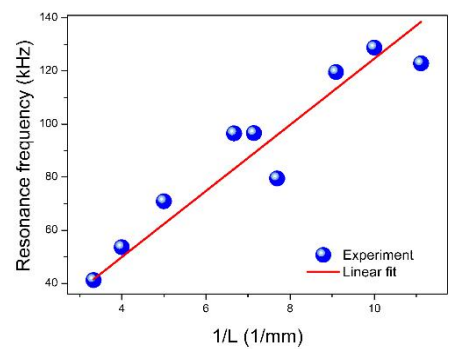

d

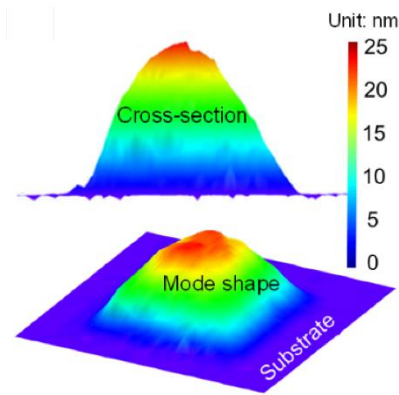

e

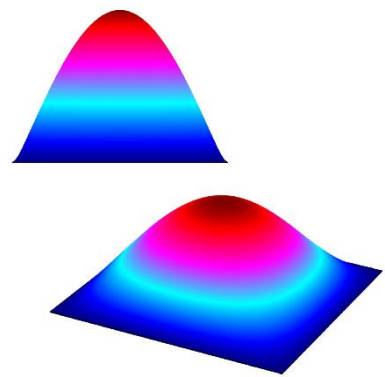

**Figure 3. Mechanical vibration mode characterization of the plasmonic super-lattice resonator.** a) Schematic representation of the experiment. b) Amplitude/phase-frequency response spectra of one plasmonic nanosheet resonator ($L = 300$ µm). c) Measured resonance frequency for several different sized plasmonic membrane resonators as a function of the side length $L$. d) Experimentally obtained fundamental mode shape (at 41.2 kHz). e) COMSOL simulation of the fundamental mode shape (42.8 kHz)

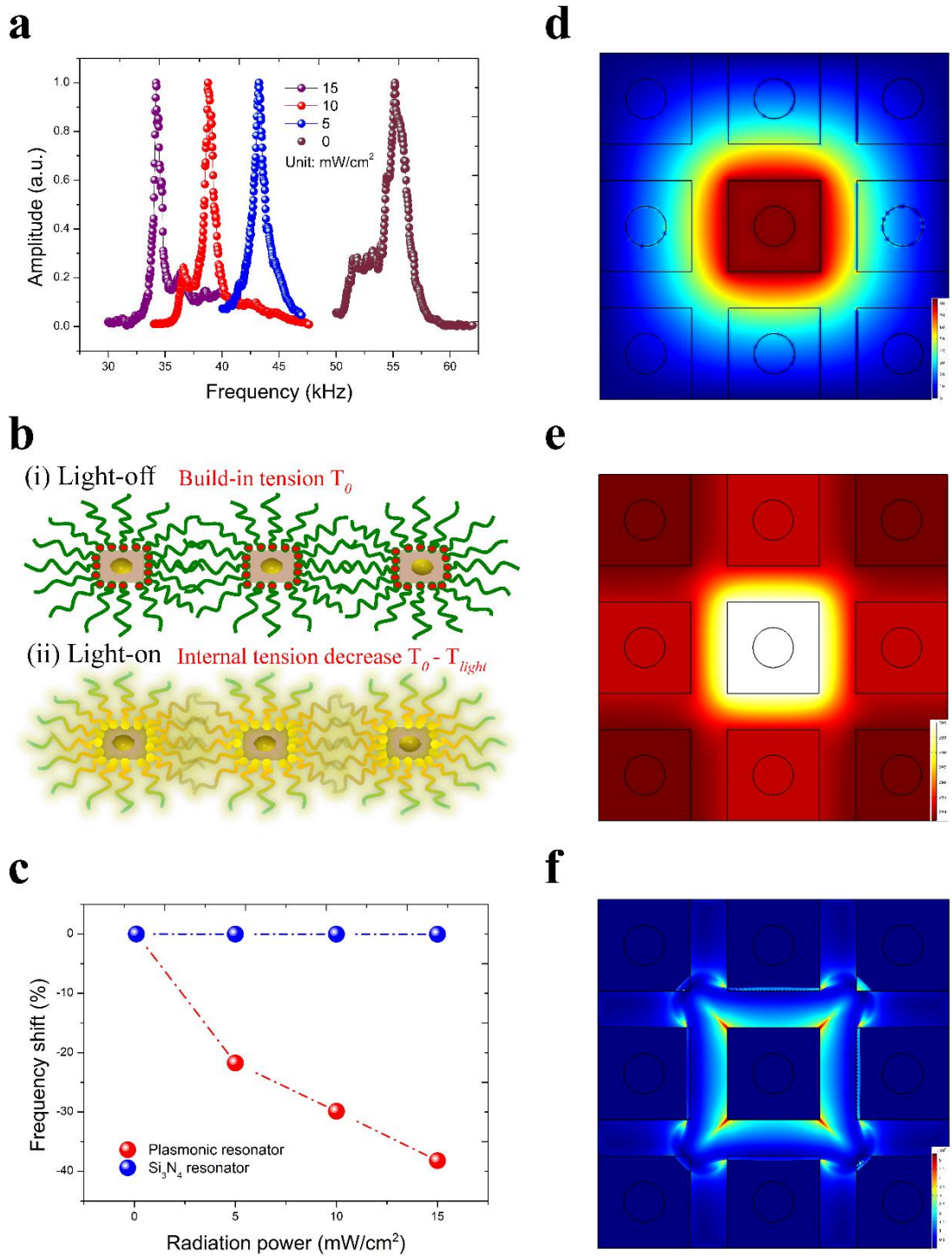

**Figure 4. Validation of the optically tunable range of the free-standing plasmonic super-lattice.** a) Frequency response curves versus radiation intensity ($L = 250$ µm). b) Physics of the photo-thermal effect inducing contraction of the polystyrene chains. There is small compression between the polystyrene links which leads to internal compression as highlighted by the gold linkages when exposed to light. c) Comparison between a plasmonic super-lattice and a $Si_3N_4$ membrane resonator. c) Electric field distribution (simulation). d) Temperature distribution (simulation). e) Stress distribution (simulation)

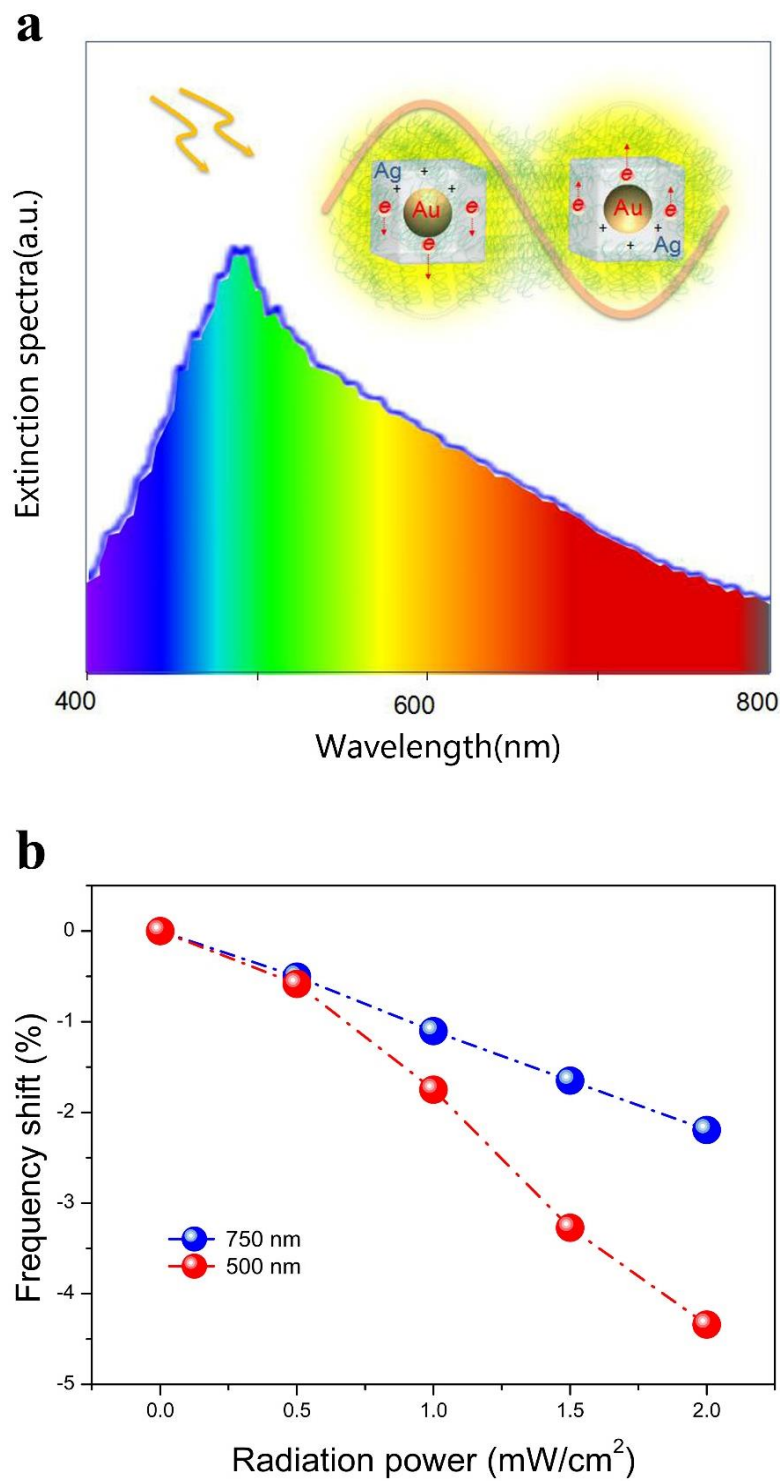

**Figure 5. Wavelength selective tuning of a plasmonic membrane resonator.** a) Experimental measured extinction spectrum of the two-dimensional plasmonic super-lattice, with a plasmonic peak resonance at ~490 nm. b) Measured resonance frequency shift under two different incident wavelengths at various incident light intensities

**The table of contents entry:**
An adaptive soft plasmonic nanosheet resonator has been fabricated based on an organic-inorganic hybrid plasmonic superlattice nanosheet that responds adaptively to either incident light intensity or wavelength. The strong light matter interaction coupling of the resonator allows for wideband tunability using low power requirements for tuning.

**Keyword** plasmonic nanosheet, light-matter interaction, optically adaptive, organic-inorganic, tunable resonance frequency

Xinghua Wang, Tanju Yildirim, Kae Jye Si, Ankur Sharma, Qinghua Qin, Qiaoliang Bao, Wenlong Cheng*, and Yuerui Lu[*]

**Title** An Adaptive Soft Plasmonic Nanosheet Resonator

ToC figure:

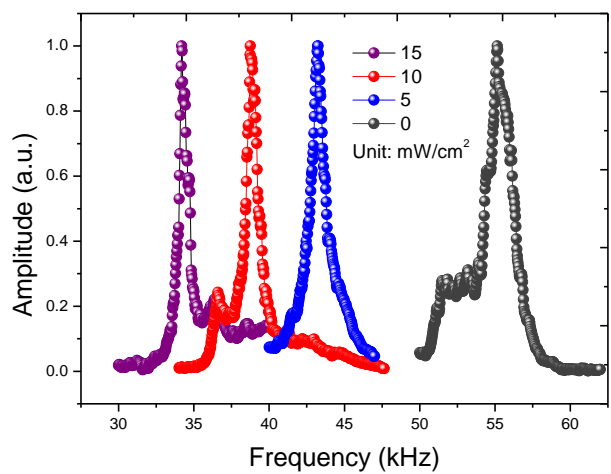

# Supporting Information

**Title: An Adaptive Soft Plasmonic Nanosheet Resonator**

Xinghua Wang,[†] Tanju Yildirim,[†] Kae Jye Si,[†] Ankur Sharma, Yunzhou Xue, Qinghua Qin, Qiaoliang Bao, Wenlong Cheng,[*] and Yuerui Lu[*]

Dr. X. Wang
Research School of Engineering, College of Engineering and Computer Science, the Australian National University, Canberra, ACT, 2601, Australia
Unmanned Systems Research Center, National Innovation Institute of Defense Technology, Academy of Military Science, Beijing, 100166, China
A. Sharma, Q Qin, Prof. Y. Lu
Research School of Engineering, College of Engineering and Computer Science, the Australian National University, Canberra, ACT, 2601, Australia
E-mail: yuerui.lu@anu.edu.au
Dr. T. Yildirim, Dr. Yunzhou Xue
College of Chemistry and Environmental Engineering, Shenzhen University, Shenzhen 518060, P. R. China
K. J. Si, Prof. W. Cheng
Department of Chemical Engineering, Faculty of Engineering, Monash University, Victoria, 3800, Australia
E-mail: wenlong.cheng@monash.edu
Prof. Q. Bao
Department of Materials Science and Engineering, ARC Centre of Excellence in Future Low-Energy Electronics Technologies (FLEET), Monash University, Clayton, Victoria, Australia
[†] These authors equally contributed to this work

**Contents**

## 1. Device Fabrication

A single-particle thick super-lattice nanosheet has been fabricated based on a drying-mediated self-assembly technique and deposited onto a silicon substrate with pre-patterned square holes. When the droplets of *Au@Ag* chloroform solution come into contact with a convex–shaped water surface, the nanoparticles at the liquid-air interface spread quickly as the chloroform evaporates as shown in Figure S1 a and b. This procedure results in the formation of a free-standing plasmonic super-lattice after water evaporation (see Figure S1c). The fabricated free-standing super-lattice nanosheet is clamped at all edges along a square hole silicon substrate with equal side lengths ($L$); the resulting device is single-particle thick in the out of plane direction as shown in Figure S1d, and the resulting large mechanical stiffness and extremely low mass make this device ideal for being used as a robust mechanical resonator in sensing and actuating applications. Inside the super-lattice nanosheet, individual metallic nanoparticles are linked to each other via the digitation of polystyrene molecular links which coat the particle surface with an innovative hybrid organic-inorganic configuration.

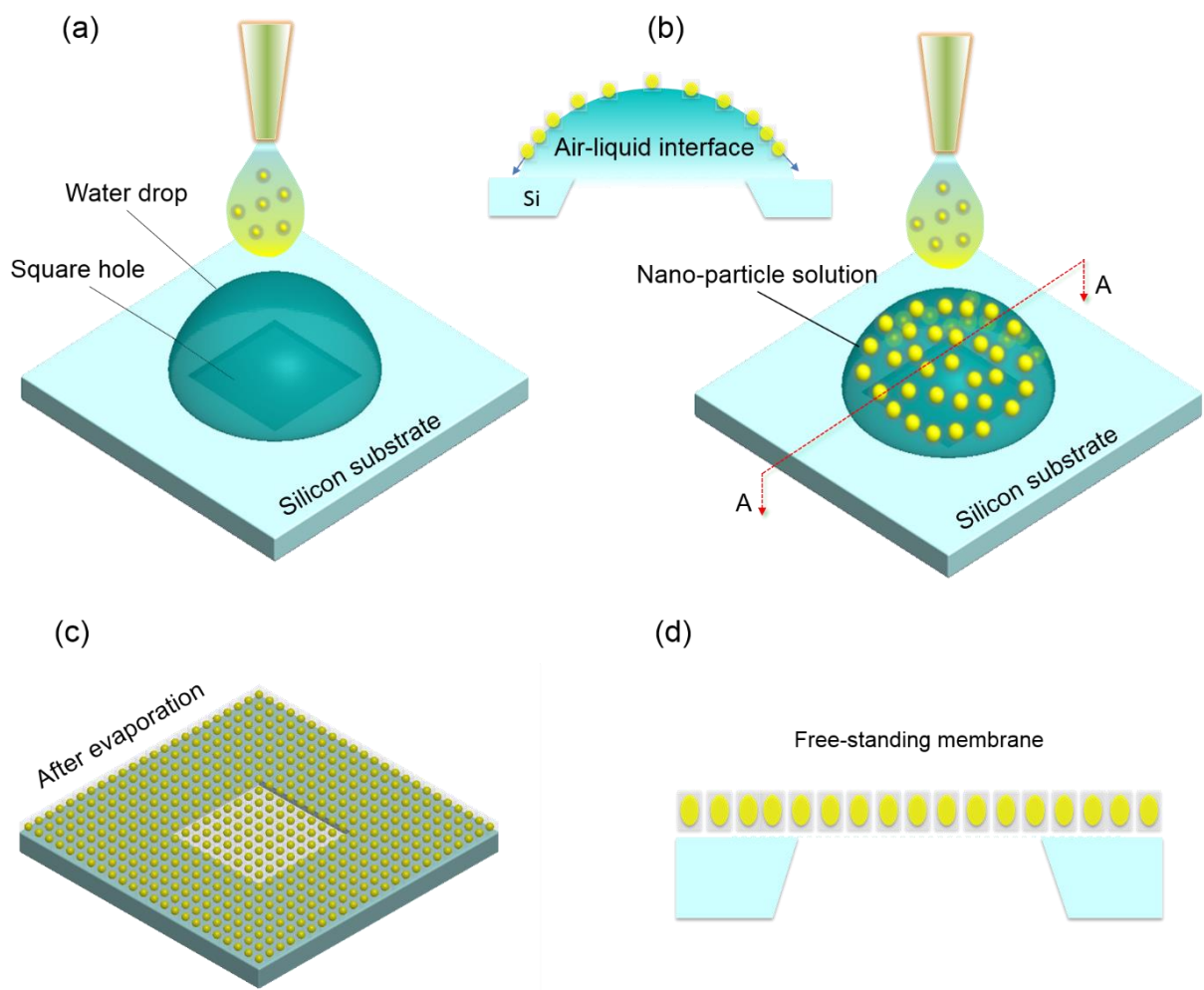

**Figure S1. Illustration of the two-dimensional free-standing plasmonic super-lattice, with well-organized Au@Ag nano-particles inter-linked by polystyrene.** a) Suspended water droplet on a silicon substrate. b) Nanoparticle polystyrene solution is introduced and spreads over the water droplet air-liquid interface. c) after water evaporation. d) Resulting 2D view of the free standing membrane.

## 2. Analytical modeling of the plasmonic resonator

In this section, a mathematical model for the temperature change internally generated by the nanoparticles inside the polystyrene matrix is formulated; moreover, a description of how this relates to the in-plane tension variation and resonance frequency shift of the plasmonic nanosheet is also discussed.

**2.1 Temperature generation inside the plasmonic resonator**

Modelling of the hybrid mechanical resonator is crucial for device design and for the interpretation of the experimental results. In the literature, very few techniques can independently control the mechanical properties of a nano-resonator. Functionalized nanoparticles are an ideal technique using the photo-thermal effect and the attached organic-polystyrene links of the plasmonic nanosheet, which can be tuned using an irradiated light source to tailor the in-plane tension of the freestanding super-lattice. Recently, the authors described the self-assembly of a hybrid organic-inorganic plasmonic membrane resonator and the mechanical dynamic response was characterised.[1] In the following derivation, the following assumption has been made, the steady-state temperature within the super-lattice is uniform and unchanging when the rate of energy of absorption is equal to the rate of heat loss; heat diffusion is a result from thermal transfer through the inorganic-organic interface via thermal conduction and convection. The overall increase of temperature within the plasmonic nanosheet originates from the release of heat energy from the optically stimulated nanoparticles and the photo-thermal effect; heat can be generated remotely by optically stimulating the nanoparticles. During the characterization process, the reflective color from the freestanding super-lattice becomes noticeable. Considering a single nanoparticle subject to a uniform light distribution, the temperature distribution generated internally in the nanoparticle and the temperature distributed to the surrounding medium is given by[2]

$$\Delta T(\mathbf{r}) = \frac{V_{NP}Q}{4\pi\kappa_s R_{NP}} \quad (when, r \leq R_{NP}) \tag{S1}$$

$$\Delta T(\mathbf{r}) = \frac{V_{NP}Q}{4\pi\kappa_s}\frac{1}{r} \quad (when, r > R_{NP})$$

where $\Delta T(r)$ is the is the temperature distribution from the nanoparticle to the outside medium, $\kappa_s$ is the thermal conductivity of the surrounding polystyrene medium, $R_{NP}$ is the radius of the nanoparticle, $V_{NP}$ is the volume of the nanoparticle and $Q$ is the heat generated inside the nanoparticle. Assuming the system is excited by an external laser field given by $I(t) = I = c\mathbf{E_0}^2 \varepsilon_0^{1/2}/8\pi$ at $t > 0$, then, $Q$, the induced electric field inside the nanoparticle is given by

$$Q = -\text{Re}\left[\frac{i\omega(\epsilon(r)-1)}{8\pi}\mathbf{E_0^2}\left|\frac{3\epsilon_0}{2\epsilon_0+\epsilon_m}\right|^2\right] \tag{S2}$$

where $\varepsilon_0$ and $\varepsilon_m$ are the dielectric constants of the surrounding medium and the metal nanoparticle, respectively. The maximum temperature generated for a single nanoparticle occurs when the radius of the nanoparticle is in contact with the surrounding medium given by

$$\Delta T_{max}(I) = \frac{R_{NP}^2}{3\kappa_s}\text{Re}\left[\frac{i\omega(\epsilon(r)-1)}{8\pi}\left|\frac{3\epsilon_0}{2\epsilon_0+\epsilon_m}\right|^2\right]\frac{8\pi I}{c\sqrt{\epsilon_0}} \tag{S3}$$

With increasing light intensity, a temperature field around the nanoparticles will form and it follows from the solution of the heat transfer equation that the function $\Delta T(r)$ decays away from the surface $\propto 1/r$. Because the distance between nanoparticles is much smaller than their size, thermal effects becomes strongly increased in nanoparticle assemblies since heat fluxes from individual nanoparticles are added. Here, the nanoparticles are connected by polystyrene polymers, which act as the molecular springs. These polystyrene organic springs are very sensitive to the environmental temperature, slight temperature differentials easily cause shifting of the polystyrene and the nanoparticles inside the matrix. During the experimental process, it was observed that very strong incident light can cause irreversible changes to the system; however, low level light intensities were reversible. The internal heating of the gold nanoparticles is due to the cumulative effect of all the nanoparticles within the plasmonic nanosheet; this means that the summation of individual heating sources should be combined over the entire device to calculate the overall internal heat generation. Inside a complex structure, the overall heat generated for an arbitrary dimensionality *m* can be estimated by [3]

$$\Delta T_{total} \approx \frac{\Delta T_{max} R_{NP}}{D} N_{NP}^{\frac{m-1}{m}} \quad (m = 2 \text{ and } 3), \tag{S4}$$

$$\Delta T_{total} \approx \frac{\Delta T_{max} R_{NP}}{D} \ln(N_{NP}) \quad (m = 1), \tag{S5}$$

where *ΔT$_{total}$* is the total temperature generated at the centre of the structure, *D* is the distance between nanoparticles (~10 *nm*) and *N$_{NP}$* is the total number of nanoparticles inside the structure.

It is important to understand the selection criteria for the matrix material selection, Figure S2 shows the maximum temperature dependence of a single gold 20 nm diameter particle under the irradiance of a 100 W/cm$^2$ light intensity, as a function of the thermal conductivity and the refractive index using Eq. S3. Gold was used as the nano particle as this was used in the experiment due to the high optical energy absorption rate. Figure S2 shows that there is a maxima achieving a 0.06 K temperature differential for a single gold nanoparticle. It is observed that the matrix material should have low thermal conductivity and a refractive index between 1.5 to 2.5 for maximum efficiency. From Table S1, it can be seen that polystyrene well suits these requirements and was selected to be the interconnecting matrix of the plasmonic nanosheet for membrane fabrication.

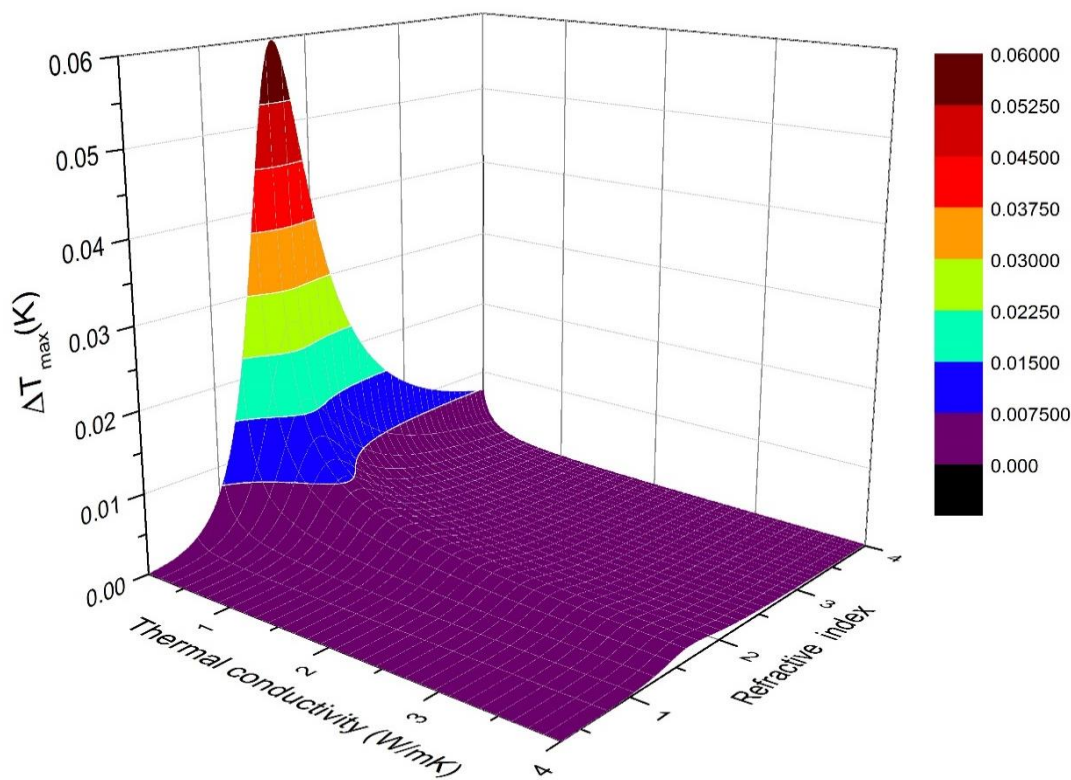

**Figure S2.** Material design selection for maximal interaction of a single gold nanoparticle with the surrounding medium.

## 2.2 Frequency shift analysis

The heat generation process within the super-lattice involves not only absorption of incident photons, but also heat transfer from the nano-particles to the surrounding polystyrene molecular links. Generally, surrounding polystyrene matrices inherently have a very small heat capacity, and consequently the internal in-plane tension of the polystyrene molecular chains rapidly compress with ambient temperature changes.[4] The unique nature of the hybrid organic-inorganic plasmonic super-lattice nanosheet has an extremely sensitive temperature-stiffness relationship. Neglecting bending stiffness, the analytical expression for the first fundamental resonance frequency of a square membrane is given by[5]

$$f_{square} = \frac{\sqrt{2}}{2L}\sqrt{\frac{T_0}{\rho t}} \tag{S6}$$

Here, the resonance frequency of the square membrane resonator is a function of the square root of the pre-tension $T_0$, membrane thickness $t$ and mass density $\rho$, and is inversely proportional to the side length ($L$) of the square membrane. Typical properties for polystyrene are given in Table S1.[6-8]

**Table S1**: Material and geometric properties of the square plasmonic nanosheet resonators.

| Name | Values |
| --- | --- |
| Pre-tension (experimental) | $0.074 \pm 0.006$ $N \cdot m^{-1}$ |
| Young's modulus | ~1 $Gpa$ |
| Mass density | $5.6 \times 10^3$ $kg \cdot m^{-3}$ |
| Thickness | $40 \pm 2$ $nm$ |
| Thermal conductivity | $0.15$ $W \cdot m^{-1} K^{-1}$ |
| Specific heat capacity | $1.25$ $J \cdot g^{-1} \cdot K^{-1}$ |
| Thermal expansion coefficient | $120 \times 10^{-6}$ $K^{-1}$ |
| Dielectric constant | 2.15 |

Specifically, the pre-tension within the membrane is induced during the fabrication process; however, due to the organic-inorganic configuration proposed in this work, the tension can be tuned by the method of heating or cooling through optical stimulation. When the polystyrene heats, the in-plane tension of the membrane decreases sharply and the induced shift of the in-plane tension is calculated as

$$\Delta T_{tens} = E\alpha t \cdot \Delta T_{temp} \tag{S7}$$

where $\Delta T_{tens}$ is the effective tension of the device, $E$ is the Young's modulus of the super-lattice and $\alpha$ is the thermal coefficient of expansion of the membrane. $\Delta T_{temp}$ is the variation of the average temperature inside the membrane. Substituting Eq. S7 into S6, the resonance frequency of the plasmonic nanosheet resonator can be formulated as

$$f_{square} = \frac{\sqrt{2}}{2L}\sqrt{\frac{T_0 - \Delta T_{Tens}}{\rho t}} = \frac{\sqrt{2}}{2L}\sqrt{\frac{T_0 - E\alpha t \cdot \Delta T_{temp}}{\rho t}} \tag{S8}$$

For a membrane-based freestanding plasmonic nanosheet, the in-plane tension could be used to tune the resonance frequency of the vibrating structure in order to target a particular ambient vibration source frequency. The wideband resonance frequency shift is induced by the expansion or contraction of polystyrene organic spring-links.

The nanoparticle matrices are tightly packed with polystyrene molecular links (inter-particle spacing is ~10 *nm*). The resonance frequencies of the freestanding super-lattice are proportional to the square root of the tension; this means that the square of the resonance frequency ($f^2$) is simply proportional to the membrane tension $T_{tens}$ such that

$$\frac{T_0}{f_0^2} = \frac{T_{tens}}{f^2} \tag{S9}$$

where $T_0$ and $f_0$ are the initial pre-tension and resonance frequency of the super-lattice, respectively. The change in the membrane tension can be written as $\Delta T_{tens} = T_{tens} - T_0$. Substituting the two equations, the ratio of the effective membrane resonance frequency to the original frequency is a function of the variation of membrane tension given by

$$\frac{f}{f_0} = \sqrt{\frac{T_{tens}}{T_0}} = \sqrt{1 + \frac{\Delta T_{tens}}{T_0}} \tag{S10}$$

To further investigate the temperature and frequency shift relationship in the plasmonic resonators, experiments were conducted for the temperature-frequency response according to Figure S3. Small changes in the temperature of the plasmonic resonator can cause large frequency shifts of almost 30% for a 10℃ temperature difference caused by light irradiation. Moreover, it was observed that the plasmonic resonators frequency shifts about 40% of its original frequency within 20℃. After this point, the resonators frequency shift becomes fairly constant with no significant changes in the resonance frequency and this may mean there is a limit to the heating of a plasmonic resonator and the frequency shift obtainable.

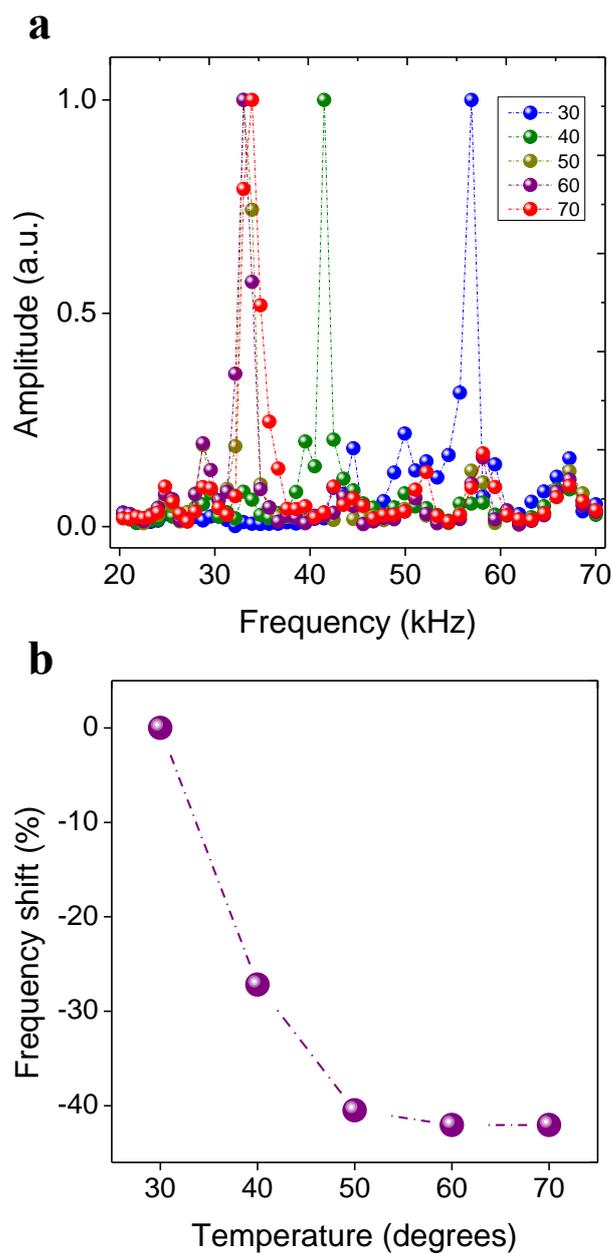

**Figure S3. Temperature dependent frequency shift of the plasmonic resonator. a)** Frequency reponse curves versus various temperatures. **b)** Frequency shift of the plasmonic resonator as a function of the temperature.

[1] K. J. Si, D. Sikdar, Y. Chen, F. Eftekhari, Z. Xu, Y. Tang, W. Xiong, P. Guo, S. Zhang, Y. Lu, Q. Bao, W. Zhu, M. Premaratne, W. Cheng, *ACS Nano* **2014**, *8*, 11086.
[2] H. H. Richardson, M. T. Carlson, P. J. Tandler, P. Hernandez, A. O. Govorov, *Nano Lett.* **2009**, *9*, 1139.
[3] A. O. Govorov, W. Zhang, T. Skeini, H. Richardson, J. Lee, N. A. Kotov, *Nanoscale Res. Lett.* **2006**, *1*, 84.
[4] M. Reismann, J. C. Bretschneider, G. v. Plessen, U. Simon, *Small* **2008**, *4*, 607.
[5] L. Dong, M. Grissom, F. T. Fisher, *AIMS Energy* **2015**, *3*, 344.
[6] S. Yu, P. Hing, X. Hu, *J. Appl. Phys.* **2000**, *88*, 398.
[7] S. Yu, P. Hing, X. Hu, *Composites Part A: Applied science and manufacturing* **2002**, *33*, 289.
[8] R. P. Sheldon, *Composite polymeric materials*, Applied Science. Elsevier Science distributor, 1982.